\documentstyle[prl,aps,floats]{revtex}

\begin{document}
\preprint{astro-ph/0003278~v2}
\draft

\input epsf

\newcommand\eq[1]{Eq.~(\ref{#1})}
\newcommand\eqs[2]{Eqs.~(\ref{#1}) and (\ref{#2})}
\newcommand\eqss[3]{Eqs.~(\ref{#1}), (\ref{#2}) and (\ref{#3})}
\newcommand\eqsss[4]{Eqs.~(\ref{#1}), (\ref{#2}), (\ref{#3})
and (\ref{#4})}
\newcommand\eqssss[5]{Eqs.~(\ref{#1}), (\ref{#2}), (\ref{#3}),
(\ref{#4}) and (\ref{#5})}
\newcommand\eqst[2]{Eqs.~(\ref{#1})--(\ref{#2})}

\newcommand\ee{\end{equation}}
\newcommand\be{\begin{equation}}
\newcommand\eea{\end{eqnarray}}
\newcommand\bea{\begin{eqnarray}}
\renewcommand{\topfraction}{0.99}
\newcommand\sub[1]{_{\rm {#1}}}

\twocolumn[\hsize\textwidth\columnwidth\hsize\csname 
@twocolumnfalse\endcsname

\title{A new approach to the evolution of cosmological perturbations
on large scales}
\author{David Wands$^1$, Karim A.~Malik$^1$, David H.~Lyth,$^2$ and
Andrew R.~Liddle$^{3,4}$}
\address{$^1$School of Computer Science and Mathematics, Mercantile House,
University of Portsmouth, Portsmouth PO1 2EG, U.~K.}
\address{$^2$Department of  Physics, University of Lancaster,
Lancaster LA1 4YB, U.~K.}
\address{$^3$Astrophysics Group, The Blackett Laboratory, 
Imperial College, Prince Consort Road, London SW7 2BZ, U.~K.} 
\address{$^4$Astronomy Centre, University of Sussex, 
Brighton BN1 9QJ, U.~K. (present address)}
\date{\today} 
\maketitle
\begin{abstract}
We discuss the evolution of linear perturbations about a
Friedmann--Robertson--Walker background metric, using only the local
conservation of energy--momentum. We show that on sufficiently large
scales the curvature perturbation on spatial hypersurfaces of
uniform-density is conserved when the non-adiabatic pressure
perturbation is negligible. This is the first time that this result
has been demonstrated independently of the gravitational field
equations.  A physical picture of long-wavelength perturbations as
being composed of separate Robertson--Walker universes gives a simple
understanding of the possible evolution of the curvature perturbation,
in particular clarifying the conditions under which super-horizon
curvature perturbations may vary.
\end{abstract}

\pacs{PACS numbers: 98.80.Cq \hfill astro-ph/0003278~v2}

\vskip2pc]

\section{Introduction}

Structure in the Universe is generally supposed to originate from the
quantum fluctuation of the inflaton field.  As each scale leaves the
horizon during inflation, the fluctuation freezes in, to become a
perturbation of the classical field. The resulting cosmological
inhomogeneity is commonly characterized by the intrinsic 
curvature of spatial hypersurfaces defined with respect to the matter.
This metric perturbation is a crucial quantity, because at approach of
horizon re-entry after inflation it determines the adiabatic
perturbations of the various components of the cosmic fluid, which
seem to give a good account of large-scale structure~\cite{LL}.

To compare the inflationary prediction for the curvature perturbation
with observation, we need to know its evolution outside the horizon,
through the end of inflation, until re-entry on each cosmologically
relevant scale.  The standard assumption is that the curvature
perturbation is practically constant. This has recently been called
into question in the context of preheating models \cite{KLS97} at the
end of inflation where non-inflaton perturbations can be resonantly
amplified \cite{Betal,LLMW}. The purpose of the present paper is to
investigate the circumstances under which the curvature perturbation
may vary.

Using only the local conservation of energy--momentum, we show that
the rate of change of the curvature perturbation on uniform-density
hypersurfaces\footnote{The ``conserved quantity'' $\zeta$ was
originally defined in Bardeen, Steinhardt and Turner \cite{BST}, but
constructed from perturbations defined in the uniform Hubble-constant
gauge.}, $\zeta$, on large scales is due to the non-adiabatic part of
the pressure perturbation.  This result is independent of the form of
the gravitational field equations, demonstrating for the first time
that the curvature perturbation remains constant on large scales for
purely adiabatic perturbations in {\em any} relativistic theory of
gravity where the energy--momentum tensor is covariantly conserved,
$T^\mu_{~\nu;\mu}=0$.  We also show that for adiabatic perturbations
produced during single field inflation the curvature perturbation on
uniform-density hypersurfaces, $\zeta$ \cite{BST,Bardeen88,MS98}, can
be identified with the comoving curvature perturbation, ${\cal R}$
\cite{LL,David+Tony}.

The pressure perturbation must be adiabatic if there is a definite
equation of state for the pressure as a function of density, which is
the case during both radiation domination and matter domination.  On
the other hand, a change in $\zeta$ on super-horizon scales will occur
during the transition from matter to radiation domination if there is
an isocurvature matter density perturbation~\cite{ks87,David+Tony}.
We give a simple derivation of this effect in terms of the curvature
perturbations on uniform-radiation and uniform-matter hypersurfaces
which remain constant throughout.

A simple intuitive understanding of how the curvature perturbation on
large scales changes, due to the different integrated expansion in
locally homogeneous but causally-disconnected regions of the universe,
can be obtained within the `separate universes' picture which we 
describe in section~\ref{sepsect}. This enables one to model the
evolution of the large-scale curvature perturbation using the
equations of motion for an unperturbed Robertson--Walker universe.
In section~\ref{sectinfl} we use this approach to discuss the
evolution of the curvature perturbation in single- and multi-field
inflation models.

\section{Linear scalar perturbations}
\label{scalpert}

In this section we summarize the essential results from cosmological
perturbation theory, applied to the scalar metric perturbations and
the associated perturbations in the pressure and energy density. 
In contrast with the usual approach to cosmological perturbation
theory, we shall not invoke any gravitational field equations. We
define energy-momentum in the usual way,
\be 
T_{\mu\nu} \equiv
-2\frac{\partial {\cal L}\sub{mat}}{\partial g^{\mu\nu}} + g_{\mu\nu}
{\cal L}\sub{mat}
 \, ,  \nonumber 
\ee 
where ${\cal L}\sub{mat}$ is any contribution to the Lagrange
density from matter fields with no external interactions.
General coordinate invariance implies the energy-momentum conservation
law $T^{\mu}_{~\nu;\mu}=0$, without invoking the Einstein field
equations.

There are many different ways of characterizing cosmological
perturbations, reflecting the arbitrariness in the choice of
coordinates (gauge), which in turn determines the slicing of spacetime
into spatial hypersurfaces, and its threading into timelike
worldlines.  The line element allowing arbitrary linear scalar
perturbations of a Friedmann--Robertson--Walker (FRW) background can
be written~\cite{lifshitz,Bardeen,KS,MFB}
\begin{eqnarray}
\label{ds}
ds^2 &=& -(1+2A) dt^2+2a^2(t) \nabla_i B\, dx^i dt 
 \nonumber \\
&& \ \  + a^2(t)
\left[(1-2\psi)\gamma_{ij}+2\nabla_i\nabla_j E \right] dx^idx^j \, .
\end{eqnarray}
The unperturbed spatial metric for a space of constant curvature
$\kappa$ is given by $\gamma_{ij}$ and covariant
derivatives with respect to this metric are denoted by $\nabla_i$.
\footnote{For comparison with the notation of
Bardeen~\cite{Bardeen} note that
\begin{eqnarray}
A \equiv A_B Q^{(0)} \, , 
&\qquad& 
\psi \equiv - \left( H_L+{1\over3}H_T \right) Q^{(0)} \ ,
\nonumber \\
B \equiv {B_B Q^{(0)} \over ka} \ , 
&\qquad& 
E \equiv {H_T Q^{(0)} \over k^2} \, ,
\end{eqnarray}
where Bardeen explicitly included $Q^{(0)}(x^i)$, the eigenmodes of
the spatial Laplacian, $\nabla^2$, with eigenvalue $-k^2$.}  The
intrinsic curvature of a spatial hypersurface, $^{(3)}R$, is usually
described by the dimensionless curvature perturbation\footnote{This
quantity is denoted ${\cal R}$ in Refs.~\cite{ss,ST}.} $\psi$, where
\begin{equation}
^{(3)}R = {6\kappa\over a^2}
 + \frac{12\kappa}{a^2} \psi + \frac{4}{a^2} \nabla^2\psi \, .
\end{equation}

The curvature perturbation on fixed-$t$ hypersurfaces
is a gauge-dependent quantity and under an
arbitrary linear coordinate transformation, $t\to t+\delta t$, it
transforms as
\begin{equation}
\label{gaugeR}
\psi \to \psi + H \delta t \, .
\end{equation}
For a scalar quantity $x$, such as the energy density or the pressure,
the corresponding transformation is
\be
\delta \rho \to \delta \rho - \dot \rho \, \delta t
\, ,
\ee
where a dot denotes differentiation with respect to coordinate
time $t$.

The curvature perturbation on uniform-density hypersurfaces, 
can be written as\footnote{The sign of $\zeta$ is chosen here to
coincide with Refs.~\cite{BST,Bardeen88}.}
\begin{equation}
\label{defzeta}
- \zeta = H \xi \ ,
\end{equation}
where the displacement between the
uniform-density ($\delta\rho=0$) hypersurface and the uniform-curvature
($\psi=0$) hypersurface has the gauge-invariant definition:
\begin{equation}
\xi \equiv {\psi \over H} + {\delta\rho \over \dot\rho} \,.
\end{equation}
Alternatively one can work in terms of the density perturbation on
uniform-curvature hypersurfaces
\begin{equation}
\label{rhoR}
\delta\rho_{\psi} = \dot\rho \, \xi \,,
\end{equation}
where the subscript $\psi$ indicates the uniform-curvature
hypersurface. 

The curvature perturbation on uniform-density hypersurfaces, $\zeta$,
is often chosen as a convenient gauge-invariant definition of the
scalar metric perturbation on large scales.
These hypersurfaces become ill-defined if the density is not strictly
decreasing, as can occur in a scalar field dominated universe when the
kinetic energy of the scalar field vanishes. In this case one can 
instead work in terms of the density perturbation on uniform-curvature
hypersurfaces, $\delta\rho_{\psi}$, which remains finite.

The pressure perturbation (in any gauge) can be split
into adiabatic and entropic (non-adiabatic) parts,
by writing 
\begin{equation}
\delta p = c_{{\rm s}}^2 \delta\rho + \dot{p} \Gamma \, ,
\end{equation}
where $c_s^2\equiv \dot p/\dot \rho$. 
The non-adiabatic part is $\delta p_{\rm nad}\equiv \dot p \Gamma$,
and 
\begin{equation}
\label{defGamma}
\Gamma \equiv {\delta p \over \dot{p}} - {\delta\rho \over \dot{\rho}}
\,.
\end{equation}
The entropy perturbation $\Gamma$, defined in this way, is
gauge-invariant, and represents the displacement between hypersurfaces
of uniform pressure and uniform density.

\section{Evolution of the curvature perturbation}
\label{curvpertsect}

\subsection{Rate of change of the curvature perturbation on large scales}

Of primary interest to us, and much of modern cosmology, is the
evolution of the curvature perturbation, $\psi$, on the constant-time
hypersurfaces defined in Eq.~(\ref{ds}).
These constant-time hypersurfaces are orthogonal to the unit time-like
vector field~\cite{KS}
\begin{equation}
\label{Nmu}
n^\mu=(1-A,-\nabla^i B) \,.
\end{equation}
The expansion  of the spatial hypersurfaces with
respect to the proper time, $d\tau\equiv(1+A)dt$, of observers with
4-velocity $n^\mu$, is given by
\begin{equation}
\label{expansion}
\theta \equiv n^\mu_{~;\mu} =3H \left(1- A \right) - 3\dot\psi +
 \nabla^2 \sigma \,,
\end{equation}
where the scalar describing the shear is
\begin{equation}
\label{shear}
\sigma = \dot{E} - B \,.
\end{equation}
However it is useful to define the expansion rate with respect to the
coordinate time
\begin{equation}
\tilde\theta = (1+A)\theta
 = 3H - 3\dot\psi + \nabla^2\sigma \,.
\end{equation}
We can write this as an equation for the time evolution of $\psi$ in
terms of the perturbed expansion,
$\delta\tilde\theta\equiv\tilde\theta-3H$, and the shear:
\begin{equation}
\label{dotpsi}
\dot\psi = -{1\over3}\delta\tilde\theta + {1\over3} \nabla^2\sigma \, .
\end{equation}
Note that this is independent of the field equations and follows
simply from the geometry.

Irrespective of the gravitational field equations we can derive
important results from the local conservation of the energy--momentum
tensor
$T^{\mu}_{~\nu;\mu}=0$.
The energy conservation equation $n^{\nu}T^\mu_{~\nu;\mu}=0$
for first-order density perturbations gives
\begin{equation}
\label{continuity}
\dot{\delta\rho} = -3H(\delta\rho + \delta p)
 + (\rho+p) \left[ 3\dot\psi - \nabla^2\left(\sigma+v+B\right) \right]
\, ,
\end{equation}
where $\nabla^i v$ is the perturbed 3-velocity of the fluid.
In the uniform-density gauge, where $\delta\rho=0$ and $\psi=-\zeta$,
the energy conservation equation~(\ref{continuity}) immediately gives
\begin{equation}
\label{dotzeta}
\dot\zeta = - {H\over \rho+p} \delta p_{\rm nad}
 - {1\over3} \nabla^2 \left(\sigma+v+B\right) \,.
\end{equation}
We emphasize that we have derived this result without invoking any
gravitational field equations, although related results have been
obtained in particular non-Einstein gravity
theories~\cite{earlierHwang,Hwang}.
We see that $\zeta$ is constant if (i) there is no non-adiabatic
pressure perturbation, and (ii) the divergence of the 3-momentum on
zero-shear hypersurfaces, $\nabla^2(v+B+\sigma)$, is
negligible.

On sufficiently large scales, gradient terms can be neglected
and~\cite{GBW,David+Tony}
\begin{equation}
\label{dotzeta2}
\dot\zeta = - {H\over \rho+p} \delta p_{\rm nad} \,,
\label{keyresult}
\end{equation}
which implies that $\zeta$ is constant if 
the pressure perturbation is adiabatic. It has been argued 
\cite{David+Tony} that the divergence is likely to be negligible
on all super-horizon scales, and in  the following we shall make
that assumption.

Although there have been many previous discussions of conserved
quantities in perturbed FRW cosmologies (which coincide with $\zeta$
on large scales), we believe that this is the first time that the 
constancy of $\zeta$ has been derived without reference to any
equations of motion for the gravitational field. It holds for
linear perturbations about an FRW metric for any relativistic theory
of gravity, as a consequence of local energy conservation
$n^\nu T^{\mu}_{~\nu;\mu}=0$.

\subsection{Non-Einstein gravity theories}

The most intensively studied example of non-Einstein gravity is
provided by scalar--tensor theories, which include a scalar field,
$\phi$, non-minimally coupled to the spacetime curvature. One approach
to studying the evolution of the metric perturbation previously
applied \cite{modes} is to perform a conformal transformation to the
Einstein frame in which the scalar field is minimally coupled to the
metric, and hence the usual Einstein gravitational field equations
hold, but non-minimally coupled to other matter fields (whose
energy--momentum tensor has non-vanishing trace). The conservation of
the total energy--momentum tensor, including the scalar field, in the
Einstein frame ensures that the curvature perturbation in this frame,
$\tilde\zeta$, will remain constant on large scales, but only so long
as $\delta \phi/\dot \phi=\delta\rho/\dot\rho$, i.e., only for
perturbations obeying the generalized adiabatic condition
$\Gamma_{\phi\rho}=0$ [see Eq.~(\ref{gengamma})], in addition to the
adiabatic condition for the fluid, $\Gamma=0$ in Eq.~(\ref{defGamma}).
However, Eq.~(\ref{dotzeta}) shows that $\zeta$ must always be
conserved on uniform density hypersurfaces in the original frame where
ordinary matter is minimally coupled, for adiabatic fluid
perturbations ($\Gamma=0$) independently of the perturbations in
$\phi$.  The two alternative definitions of the curvature perturbation
are equal, $\zeta=\tilde\zeta$, only in the special case when $\delta
\phi/\dot
\phi=\delta\rho/\dot\rho$ and it then follows that the curvature
perturbation is constant in both frames because the generalized
adiabatic condition holds.

Non-Einstein gravity (in our four spacetime dimensions) may also
emerge \cite{SMS99} from theories involving a large extra dimension
\cite{hw,RS99}. In particular, our proof of \eq{keyresult} validates a
recent discussion \cite{MWBH99} of chaotic inflation in these
theories, which relied on that equation.

\subsection{Matter plus radiation}

In a multi-fluid system we can define
uniform-density hypersurfaces for each fluid and a corresponding
curvature perturbation on these hypersurfaces, $\zeta_{(i)}\equiv
-\psi-\delta\rho_{(i)}/\dot\rho_{(i)}$.  Equation~(\ref{dotzeta}) then
shows that $\zeta_{(i)}$ {\em remains constant for adiabatic perturbations
in any fluid whose energy--momentum is locally conserved:} $n^\nu
T^{~\mu}_{(i)~\nu;\mu}=0$.  Thus, for example, in a universe
containing non-interacting cold dark matter plus radiation, which both
have well-defined equations of state ($p_{\rm{m}}=0$ and
$p_\gamma=\rho_\gamma/3$), the curvatures of uniform-matter-density
hypersurfaces, $\zeta_m$, and of uniform-radiation-density
hypersurfaces, $\zeta_\gamma$, remain constant on super-horizon
scales. The curvature perturbation on the uniform-total-density
hypersurfaces is given by
\begin{equation}
\zeta = \frac{(4/3)\rho_\gamma\zeta_\gamma + \rho_m\zeta_{\rm{m}}}
{(4/3)\rho_\gamma + \rho_m} \,.
\label{matplusrad}
\end{equation}
At early times in the radiation dominated era 
($\rho_\gamma\gg\rho_{\rm{m}}$)
we have $\zeta_{\rm{init}}\simeq\zeta_\gamma$, while at late times
($\rho_{\rm{m}}\gg \rho_\gamma$) we have 
$\zeta_{\rm{fin}}\simeq\zeta_{\rm{m}}$. 
$\zeta$ remains constant throughout only for adiabatic perturbations 
where the uniform-matter-density and uniform-radiation-density 
hypersurfaces coincide, ensuring $\zeta_\gamma=\zeta_{\rm{m}}$. 
The isocurvature (or entropy) perturbation is conventionally 
denoted by the perturbation in
the ratio of the photon and matter number densities
\begin{equation}
S 
= {\delta n_\gamma \over n_\gamma} 
- {\delta n_{\rm{m}} \over n_{\rm{m}}} =
3 \left( \zeta_\gamma - \zeta_{\rm{m}} \right) \,.
\end{equation}
Hence the entropy perturbation for any two non-interacting fluids
always remains constant on large scales independent of the 
gravitational field equations.
Hence we recover the standard result for the final curvature
perturbation in terms of the initial curvature and entropy
perturbation\footnote
{This result was  derived first  by solving a differential 
equation \cite{ks87}, and then  \cite{David+Tony} by integrating 
\eq{keyresult} using \eq{matplusrad}. We have here demonstrated
that even the integration is unnecessary.}
\begin{equation}
\zeta_{\rm{fin}} = \zeta_{\rm{ini}} - {1\over3} S \,.
\end{equation}

\section{The separate universe approach}
\label{sepsect}

One can proceed to use the perturbed field equations, to follow the
evolution of linear perturbations in the metric and matter fields in
whatever gauge one chooses.  This allows one to calculate the
corresponding perturbations in the density and pressure and the
non-adiabatic pressure perturbation if there is one, and see whether
it causes a significant change in $\zeta$.

However, there is a particularly simple alternative approach to
studying the evolution of perturbations on large scales, which has
been employed in some multi-component inflation models
\cite{star,salo,ss,ns,ST,David+Tony}.  This considers each
super-horizon sized region of the Universe to be evolving like a
separate Robertson--Walker universe where density and pressure may take
different values, but are locally homogeneous.  After patching
together the different regions, this can be used to follow the
evolution of the curvature perturbation with time.  Figure~1 shows the
general idea of the separate universe picture, though really every
point is viewed as having its own Robertson--Walker region surrounding
it.

\begin{figure}[t]
\centering 
\leavevmode\epsfysize=9cm \epsfbox{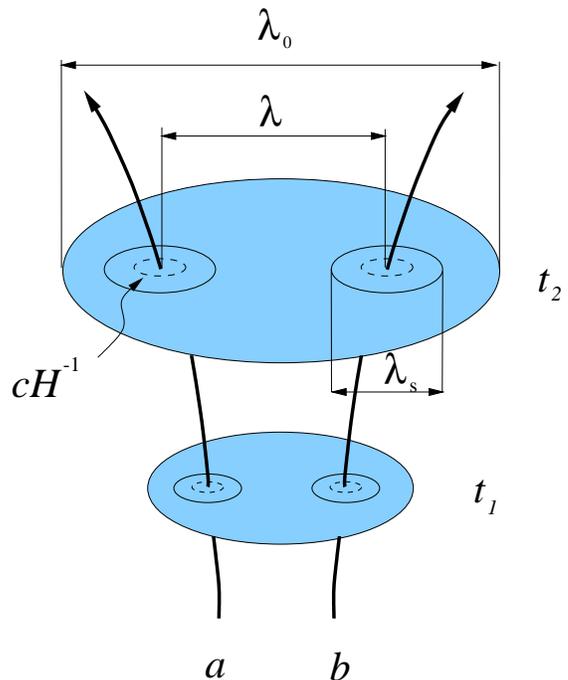}\\
\caption[sepfig]{\label{sepfig} A schematic illustration of the separate 
universes picture, with the symbols as identified in the text.}
\end{figure} 

Consider two such locally homogeneous regions $(a)$ and $(b)$ at fixed
spatial coordinates, separated by a coordinate distance $\lambda$, on
an initial hypersurface (e.g., uniform-density hypersurface) specified
by a fixed coordinate time, $t=t_1$, in the appropriate gauge (e.g.,
uniform-density gauge). The initial large-scale curvature perturbation
on the scale $\lambda$ can then be defined (independently of the
background) as
\begin{equation}
\delta\psi_1\equiv\psi_{a1}-\psi_{b1} \,.
\end{equation}
On a subsequent hypersurface defined by $t=t_2$ the curvature
perturbation at $(a)$ or $(b)$ can be evaluated using
Eq.~(\ref{dotpsi}) [but neglecting $\nabla^2\sigma$] to give~\cite{ss}
\begin{equation}
\psi_{a2} = \psi_{a1} - \delta N_a \,,
\end{equation}
where the integrated expansion between the two hypersurfaces along the
world-line followed by region $(a)$ is given by
$N_a=N+\delta N_a$, with $N\equiv\ln a$ the expansion in the
unperturbed background and
\begin{equation}
\delta N_a = \int_1^2 {1\over3} \delta\tilde\theta_a dt \,.
\end{equation}
The curvature perturbation when $t=t_2$ on the comoving scale
$\lambda$ is thus given by
\begin{equation}
\label{dpsi2}
\delta\psi_2 \equiv \psi_{a2}-\psi_{b2}
 = \delta\psi_1 - \left( N_a - N_b \right) \,.
\label{neq}
\end{equation}
In order to calculate the change in the curvature perturbation in any
gauge on very large scales it is thus sufficient to evaluate the
difference in the integrated expansion between the initial and final
hypersurface along different world-lines.

In particular, using \eq{neq}, one can evolve the curvature
perturbation, $\zeta$, on super-horizon scales, knowing only the
evolution of the family of Robertson--Walker universes, which according
to the separate Universe assumption describe the evolution of the
Universe on super-horizon scales:
\begin{equation}
\Delta\zeta = \Delta N \,,
\end{equation}
where $\Delta\zeta=-\psi_a+\psi_b$ on uniform-density hypersurfaces and
$\Delta N=N_a-N_b$ in Eq.~(\ref{dpsi2}). 
As we shall discuss in the next section, this evolution is in turn
specified by the values of the relevant fields during inflation, and
as a result one can calculate $\zeta$ at horizon re-entry from the
vacuum fluctuations of these fields.

While it is a non-trivial assumption to suppose that every comoving
region {\em well outside the horizon} evolves like an unperturbed
universe, there has to be some scale $\lambda_{\rm s}$ for which that
assumption is true to useful accuracy.  If there were not, the concept
of an unperturbed (Robertson--Walker) background would make no sense.
We use the phrase `background' to describe the evolution on a
much larger scale $\lambda_0$, which should be much bigger even than
our present horizon size, with respect to which the perturbations in
section~\ref{scalpert} were defined.
It is important to distinguish this from regions of size $\lambda_{\rm s}$
large enough to be treated as locally homogeneous, but which when
pieced together over a larger scale, $\lambda$, represent the
long-wavelength perturbations under consideration.  
Thus we require a hierarchy of scales:
\begin{equation}
\lambda_0 \gg \lambda \gg \lambda_{{\rm s}} \gtrsim cH^{-1}\ .  
\end{equation} 
Ideally $\lambda_0$ would be taken to be infinite. However it may be
that the Universe becomes highly inhomogeneous on some very much
larger scale, $\lambda_{{\rm e}} \gg\lambda_0$, where effects such as
stochastic or eternal inflation determine the dynamical evolution.
Nevertheless, this will not prevent us from defining an effectively
homogeneous background in our observable Universe, which is governed
by the local Einstein equations and hence impervious to anything
happening on vast scales.  Specifically we will assume that it is
possible to foliate spacetime on this large scale $\lambda_0$ with
spatial hypersurfaces.

When we use homogeneous equations to describe separate regions on
length scales greater than $\lambda_{\rm s}$, we are implicitly assuming
that the evolution on these scales is independent of shorter
wavelength perturbations. This is true within linear perturbation
theory in which the evolution of each Fourier mode can be considered
independently, but any non-linear interaction introduces mode-mode
coupling which undermines the separate universes picture.  
The separate universe model may still be used for the evolution of
linear metric perturbation if the perturbations in the total density
and pressure remain small, but a suitable model (possibly a
thermodynamic description) of the effect of the non-linear evolution
of matter fields on smaller scales may be necessary in some cases.  An
application to the study of preheating at the end of inflation is
discussed in Section~\ref{sectpreheat}.

Adiabatic perturbations in the density and pressure correspond to
shifts forwards or backwards in time along the background
solution, $\delta p/\delta\rho=\dot{p}/\dot\rho\equiv c_s^2$, and
hence $\Gamma=0$ in Eq.~(\ref{defGamma}). 
For example, in a universe containing only baryonic matter plus
radiation, the density of baryons or photons may vary locally, but the
perturbations are adiabatic if the ratio of photons to
baryons remains unperturbed.
Different regions are compelled to undergo the same evolution along a
unique trajectory in field space, separated only by a shift in the
expansion.  
The pressure $p$ thus remains a unique function of the density $\rho$
and the energy conservation equation, $d\rho/dN=-3(\rho+p)$,
determines $\rho$ as a function of the integrated expansion, $N$.
Under these conditions, uniform-density hypersurfaces are separated by
a uniform expansion and hence the curvature perturbation, $\zeta$,
remains constant.

For $\Gamma\neq0$ it is no longer possible to define a simple shift to
describe both the density and pressure perturbation.  The existence of
a non-zero pressure perturbation on uniform-density hypersurfaces
changes the equation of state in different regions of the Universe and
hence leads to perturbations in the expansion along different
worldlines between uniform-density hypersurfaces.
This is consistent with Eq.~(\ref{dotzeta}) which quantifies how the
non-adiabatic pressure perturbation determines the variation of
$\zeta$ on large scales \cite{GBW,David+Tony}.

The entropy perturbation between any two quantities (which are
spatially homogeneous in the background) has a naturally
gauge-invariant definition 
[which follows from the obvious extension of Eq.~(\ref{defGamma})]
\begin{equation}
\label{gengamma}
\Gamma_{xy} \equiv {\delta x \over \dot{x}} - {\delta y \over \dot{y}}
\,.
\end{equation}
We define a generalized adiabatic condition which requires
$\Gamma_{xy}=0$ for any physical scalars $x$ and $y$. In the separate
universes picture this condition ensures that if all field
perturbations are adiabatic at any one time (i.e. on any spatial
hypersurface), then they must remain so at any subsequent time. 
Purely adiabatic perturbations can never give rise to entropy
perturbations on large scales as all fields share the same time shift,
$\delta t = \delta x/\dot{x}$, along a single phase-space trajectory.

\section{Inflation}
\label{sectinfl}

\subsection{Single-component inflaton field}

In Section~\ref{curvpertsect} we showed that the curvature
perturbation $\zeta$ on the uniform-density gauge is constant on large
scales for adiabatic perturbations.  A common application of this is
to perturbations produced by a single scalar field during inflation. 
Even this apparently simple case is somewhat subtle since a
scalar field obeys a second-order equation of motion and cannot in
general be described by an equation of state $p(\rho)$, since the
total energy can be split between potential and kinetic energy.
However, the existence of an attractor solution for a strongly-damped
inflaton field allows one to drop the decaying mode as inflation
progresses and ensures a unique relation between the field value and
its first derivative.

The specific relations between the inflaton field and curvature
perturbations depends on the choice of gauge.  In practice the
inflaton field perturbation spectrum can be calculated on
uniform-curvature ($\psi=0$) slices, where the field perturbations
have the gauge-invariant definition~\cite{Mukhanov,MFB}
\begin{equation}
\delta \phi_\psi \equiv \delta \phi + {\dot \phi\over H} \psi \,.
\end{equation}
In the slow-roll limit the amplitude of field fluctuations at horizon
crossing ($\lambda=H^{-1}$) is given by $H/2\pi$. 
Note that this is the amplitude of the asymptotic solution on large
scales.
This result is independent of the geometry and holds for a massless
scalar field in de Sitter spacetime independently of the gravitational
field equations.

The field fluctuation is then related to the curvature perturbation
on comoving hypersurfaces (on which the scalar field is uniform,
$\delta\phi_c=0$) using Eq.~(\ref{gaugeR}), by
\begin{equation}
{\cal R} \equiv \psi_c = 
 \frac{H}{\dot{\phi}} \delta \phi_{\psi} \, .
\end{equation}

We will now demonstrate that for adiabatic perturbations we can
identify the curvature perturbation on comoving hypersurfaces, ${\cal
R}$, with the curvature perturbation on uniform-density hypersurfaces,
$-\zeta$.  
In an arbitrary gauge the density and pressure perturbations
of a scalar field are given by
\begin{eqnarray}
\delta \rho &=& \dot\phi\,\dot{\delta\phi} - A \dot\phi^2 + V' \, \delta\phi
\,,\\
\delta p &=& \dot\phi\,\dot{\delta\phi} - A \dot\phi^2   - V' \, \delta\phi
\,,
\end{eqnarray}
where $V'\equiv dV/d\phi$.  Thus we find $\delta\rho-\delta
p=2V'\delta\phi$. For adiabatic perturbations on uniform-density
hypersurfaces both the density and pressure perturbation must vanish
and thus so does the field perturbation $\delta\phi_\rho=0$ for
$V'\neq0$. Hence the uniform-density and comoving hypersurfaces
coincide, and ${\cal R}$ and $-\zeta$ are identical, for adiabatic
perturbations.


The asymptotic solution/growing mode for the scalar field vacuum
fluctuation corresponds to a perturbation about the 
background attractor solution and hence generates a purely adiabatic
perturbation on super-horizon scales.
Thus the density perturbation when a mode re-enters the horizon during
the radiation or matter dominated eras can be directly related to 
the growing mode of the inflaton field perturbation
when that mode left the horizon during inflation
due to the constancy of $\zeta$ 
once the decaying mode becomes negligible after horizon
crossing~\cite{MS98}.
We have shown that this does not depend on any slow-roll
type approximation for the inflaton field, nor does it depend on the
form of the gravitational field equations. The result holds for any
metric theory of gravity that respects local conservation of
energy--momentum. As an example, the large-scale curvature perturbation
spectrum produced during a period of ``brane inflation'' has recently
been calculated~\cite{MWBH99} in the four-dimensional effective theory
of gravity induced on the world-volume of a 3-brane in
five-dimensional Einstein gravity~\cite{RS99,SMS99}, even though the
full theory of cosmological perturbations has yet to be determined in
this model.

\subsection{Multi-component inflaton field}

During a period of inflation it is important to distinguish between
``light'' fields, whose effective mass is less than the Hubble
parameter, and ``heavy'' fields whose mass is greater than the Hubble
parameter. Long-wavelength (super-Hubble scale) perturbations of heavy
fields are under-damped and oscillate with rapidly decaying amplitude
($\langle\phi^2\rangle\propto a^{-3}$) about their vacuum
expectation value as the universe expands. Light fields, on the other
hand, are over-damped and may decay only slowly towards the minimum of
their effective potential. It is the slow-rolling of these light
fields that controls the cosmological dynamics during inflation.

The inflaton, defined as the direction of the classical evolution, is
one of the light fields, while the other light fields (if any) will be
taken to be orthogonal to it in field space.
In a multi-component inflation model there is a family of inflaton
trajectories, and the effect of the orthogonal perturbations is to
shift the inflaton from one trajectory to another.

If all the fields orthogonal to the inflaton are heavy then there is a
unique inflaton trajectory in field space. In this case even a curved
path in field space, after canonically normalizing the inflaton
trajectory, is indistinguishable from the case of a straight
trajectory, and leads to no variation in $\zeta$.

When there are multiple light fields evolving during inflation,
uncorrelated perturbations in more than one field will lead to
different regions that are not simply time translations of each
other. In order to specify the evolution of each locally homogeneous
universe one needs as initial data the value of every cosmologically
significant field. In general, therefore, there will be non-adiabatic
perturbations, $\Gamma_{xy}\neq0$.

If the local integrated expansion, $N$, is sensitive to the
value of more than one of the light fields then $\zeta$ is able to
evolve on super-horizon scales, as has been shown by several authors
\cite{modes,GBW}.  Note also that the comoving and uniform-density
hypersurfaces need no longer coincide in the presence of non-adiabatic
pressure perturbations.  In practice it is necessary to follow the
evolution of the perturbations on super-horizon scales in order to
calculate the curvature perturbation at later times.
In most models studied so far, the trajectories converge to a unique
one before the end of inflation, but that need not be the case in
general.

The separate universe approach described in section~\ref{sepsect}
gives a rather straightforward procedure for calculating the evolution
of the curvature perturbation, $\psi$, on large scales based on the
change in the integrated expansion, $N$, in different locally
homogeneous regions of the universe. This approach was developed in
Refs.~\cite{ss,ns,ST} for general relativistic models where scalar
fields dominate the energy density and pressure, though it has not
been applied to many specific models.  In the case of a
single-component inflaton, this means that on each comoving scale,
$\lambda$, the curvature perturbation, $\zeta$, on uniform-density (or
comoving) hypersurfaces {\em must} stop changing when gradient terms
can be neglected ($\lambda>\lambda_s$). More generally, with a
multi-component inflaton, the perturbations generated in the fields
during inflation will still determine the curvature perturbation,
$\zeta$, on large scales, but one needs to follow the time evolution
during the entire period a scale remains outside the horizon in order
to evaluate $\zeta$ at later times. This will certainly require
knowledge of the gravitational field equations and may also involve
the use of approximations such as the slow-roll approximation to
obtain analytic results.

\subsection{Preheating}
\label{sectpreheat}

During inflation, every field is supposed to be in the vacuum state
well before horizon exit, corresponding to the absence of particles.
The vacuum fluctuation cannot play a role in cosmology unless it is
converted into a classical perturbation, defined as a quantity which
can have a well-defined value on a sufficiently long time-scale
\cite{guthpi,Lyth85}.  For every light field this conversion occurs at
horizon exit ($\lambda\sim H^{-1}$).
In contrast, heavy fields become classical, if at all, only when their
quantum fluctuation is amplified by some other mechanism.

There has recently been great interest in models where vacuum
fluctuations become classical (i.e., particle production occurs) due
to the rapid change in the effective mass (and hence the vacuum state)
of one or more fields. This usually (though not always~\cite{Rocky})
occurs at the end of inflation when the inflaton oscillates about its
vacuum expectation value which can lead to parametric amplification of
the perturbations --- a process which has become known as
preheating~\cite{KLS97}.  The rate of amplification tends to be
greatest for long-wavelength modes and this has lead to the claim that
rapid amplification of non-adiabatic perturbations could change the
curvature perturbation, $\zeta$, even on very large
scales~\cite{Betal}.

Within the separate universes picture this is certainly possible if
preheating leads to different integrated expansion in different
regions of the universe. In particular $\zeta$ can evolve if a
significant non-adiabatic pressure perturbation is produced on large
scales. However it is also apparent in the separate universes picture
that no non-adiabatic perturbation can subsequently be introduced on
large scales if the original perturbations were purely adiabatic.
This is of course also apparent in the field equations where
preheating can only amplify pre-existing field fluctuations. 

Efficient preheating requires strong coupling between the inflaton and
preheating fields which typically leads to the preheating field being
heavy during inflation (when the inflaton field is large).  The strong
suppression of super-horizon scale fluctuations in heavy fields during
inflation means that in this case no significant change in $\zeta$ is
produced on super-horizon scales before back-reaction due to particle
production on much smaller scales damps the oscillation of the
inflaton and brings preheating to an end~\cite{jedam,ivan,LLMW}.

Because the first-order effect is so strongly suppressed in such
models, the dominant effect actually comes from second-order
perturbations in the fields~\cite{jedam,ivan,LLMW}. 
The expansion on large scales is no longer independent of shorter
wavelength field perturbations when we consider higher-order terms in
the equations of motion.
Nonetheless in many cases it is still possible to use linear
perturbation theory for the metric perturbations while including
second-order perturbations in the matter fields.\footnote{Formally one
considers  
the matter field perturbations to be
of order $\epsilon$, but the metric perturbations to be of order
$\epsilon^2$.}
In Ref.~\cite{LLMW} this was done to show that even allowing for
second-order field perturbations, there is no significant
non-adiabatic pressure perturbation, and hence no change in $\zeta$,
on large scales in the original model of preheating in chaotic
inflation.

More recently a modified version of preheating has been
proposed~\cite{Betal2} (requiring a different model of inflation)
where the preheating field is light during inflation, and the coupling
to the inflaton only becomes strong at the end of inflation. In such a
multi-component inflation model non-adiabatic perturbations are no
longer suppressed on super-horizon scales and it is possible for the
curvature perturbation $\zeta$ to evolve both during inflation and
preheating, as described in Section~V-B.

\section{Conclusions}

In this paper, we have identified the general condition under which
the super-horizon curvature perturbation on spatial hypersurfaces can
vary as being due to differences in the integrated expansion along
different worldlines between hypersurfaces.  As long as linear
perturbation theory is valid, then, when spatial gradients of the
perturbations are negligible, such a situation can be described using
the separate universes picture, where regions are evolved according to
the homogeneous equations of motion. 

In particular, the curvature perturbation on uniform-density
hypersurfaces, $\zeta$, can vary only in the presence of a significant
non-adiabatic pressure perturbation.  The result follows directly from
the local conservation of energy--momentum and is independent of the
gravitational field equations. Thus $\zeta$ is conserved for adiabatic
perturbations on sufficiently large scales in any metric theory of
gravity, including scalar--tensor theories of gravity or induced
four-dimensional gravity in the brane-world scenario.

Multi-component inflaton models are an example where non-adiabatic
perturbations may cause the curvature perturbation to evolve on
super-horizon scales.

\section*{Acknowledgments}

We thank Marco Bruni and Roy Maartens for useful discussions. DW is
supported by the Royal Society.

 
\end{document}